\documentstyle[preprint,prd,epsf,graphics,aps,floats]{revtex}
\draft
\tightenlines

\begin{document}

\newcommand{\inclfig}[2]{
\mbox{\epsfxsize=#1cm \epsfbox{#2.eps}}
}

\preprint{hep-ph/9812490}

\title{Scheme dependence of NLO corrections to exclusive processes}
\author{
        D. M\"uller
        }
\address{Institut f\"ur Theoretische Physik,
			Universit\"at Regensburg, 93040 Regensburg, Germany
        }

\date{15.03.1999}
\maketitle

\begin{abstract}
We apply the so-called conformal subtraction scheme to predict
perturbatively exclusive processes beyond leading order. Taking into
account evolution effects, we study the scheme dependence for the
photon-to-pion transition form factor and the electromagnetic pion form
factor at next-to-leading order for different pion distribution amplitudes.
Relying on the conformally covariant operator product expansion and using
the known higher order results for polarized deep inelastic scattering,
we are able to predict perturbative corrections to the hard-scattering
amplitude of the photon-to-pion transition form factor beyond
next-to-leading order in the conformal scheme restricted to the conformal
limit of the theory.
\end{abstract}

\vspace{7cm}

\noindent
Keywords: 

{\ } \hspace{0.7cm} exclusive processes, conformal symmetry,
perturbative corrections, 

{\ } \hspace{0.7cm} scheme dependence

\noindent
PACS numbers:
\pacs{11.10.Hi, 12.38.Bx, 13.40.Gp, 13.60.Le}

\newpage
\narrowtext

\section{Introduction}

An important theoretical issue of the perturbative QCD approach to exclusive
processes is to check its self-consistency by studying perturbative
corrections. This task was carried out in the modified minimal subtraction
\mbox{ $\overline{\rm MS} $ } scheme for the next-to-leading (NLO)
corrections to the photon-to-pion transition form factor Ref.\
\cite{AguCha81,Bra83,KadMikRad86}, the electromagnetic pion form factor
\cite{FieGupOttCha81,DitRad81,Dit82,Sar82,RadKha85,BraTse87,MelNizPas98},
and the two-photon process $\gamma\gamma\to M^+ M^- (M=\pi,K)$ \cite{Niz87}.
The latter was only calculated for the special case of an equal momentum
sharing distribution amplitude (DA). Only in a recent analysis of the
electromagnetic pion form factor \cite{MelNizPas98} has the NLO correction
to the evolution of the DA been completely taken into account. Discrepancies
in the one-loop approximation of the hard-scattering amplitude for the pion
form factor were clarified in Refs.\ \cite{RadKha85,BraTse87}. The NLO
corrections to the pion form factor and to the process $\gamma\gamma\to M^+
M^-$ are rather large at accessible momentum transfer; however, it turned
out that for the equal momentum sharing DA the perturbative corrections to
the integrated cross section of the charged meson production can almost be
absorbed into the electromagnetic meson form factor (see Ref.\
{\cite{Niz87}}).

It was mentioned long ago that conformal symmetry plays an important role
for the perturbative description of exclusive processes
\cite{BroFriLepSac80,EfrRad80a}. Unfortunately, conformal symmetry is
broken by the renormalization of the UV-divergencies appearing in the
hard-scattering part as well as in the DA. There are two different sources
for the breaking of conformal symmetry induced by: (i) the running coupling
and (ii) the renormalization of the DA. While the first one is proportional
to the $\beta$-function and vanishes for a non-perturbative hypothetical
fixed point, the latter appears in the minimal subtraction (MS) scheme;
fortunately, it is absent in a special subtraction scheme, which we will
call conformal subtraction (CS) scheme \cite{Mue97a}. Let us point out that
the Brodsky-Lepage-Mackenzie (BLM) scale setting prescription is suitable to
absorb conformal anomalies, which are in the CS scheme only proportional to
$\beta$, in the scale setting prescription of the coupling.

The understanding of conformal symmetry and its breaking in perturbative QCD
has been applied recently for the NLO calculation of the leading twist-2
Wilson-coefficients \cite{BelMue97a} and the flavor singlet evolution
kernels \cite{BelMue98a,BelMue98c} relevant for off-forward processes as
well as to prove the hypotheses of naive non-Abelianzation for the
renormalon chain appearing in the evolution kernel \cite{BelMue97a}. In this
paper we pursue the phenomenological consequences of the CS scheme for the
exclusive processes mentioned above.

The rest of the paper is organized as follow. In Section \ref{scheme} we
discuss the scheme dependence of the perturbative corrections and argue that
the NLO corrections calculated in the \mbox{ $\overline{\rm MS} $ } scheme
should be reanalyzed in this special scheme that ensures conformal symmetry.
The consequences of the BLM scale fixing prescription and the evolution
effects of the pseudo-scalar meson DA to NLO are analyzed in the flavor
nonsinglet sector in Section \ref{NLO-DA}. In Sections \ref{TFF} and
\ref{FF} we numerically investigate the NLO corrections for the
photon-to-pion transition form factor and the elastic form factor,
respectively, where we study quite different parametrizations of the pion
DA. Applying the conformal operator product expansion, which is valid in the
CS scheme, we are able to predict the photon-to-pion transition form factor
to next-to-next-to-leading order (NNLO) restricted to the conformal limit of
the theory. At the moment evolution effects can not be taken into account at
this order. Conclusions are given in Section \ref{Concl}. The decomposition
and the conformal partial wave expansion of functions appearing in the
hard-scattering part of the electromagnetic form factor are listed in two
Appendices.

\section{Scheme dependence of pQCD predictions}
\label{scheme}

In Ref.\ \cite{Mue97a} we introduced the conformally covariant subtraction
(CS) scheme, which ensures that the underlying conformal structure appearing
in exclusive processes is manifest in the conformal limit. Calculations of
hard-scattering amplitudes and evolution kernels are carried out usually in
the $\overline{\mbox{MS}}$ scheme. Both schemes are related to each other by
a (finite) refactorization. For instance, the photon-to-pion transition form
factor $F_{\gamma\pi}$ at large momentum transfer factorizes in both schemes
according to the standard factorization scheme (SFA)
\cite{BroLep80,BroLep81}:
\begin{eqnarray}
\label{RG-transfor}
F_{\gamma\pi}(\omega,Q) &=&
	T^{\rm MS}(\omega,x,Q,\mu)\otimes\phi^{\rm MS}(x,\mu)
\\
	&=&
	T^{\rm CS}(\omega,x,Q,\mu)\otimes\phi^{\rm CS}(x,\mu),
\qquad 	\otimes \equiv \int_0^1 dx,
\nonumber
\end{eqnarray}
where the hard-scattering part $T(\omega,x,Q,\mu)$, depending on the
kinematical variables $\omega$ and $Q$, the momentum fraction $x$ as well as
the factorization scale $\mu$, can be calculated in
perturbation theory. The DA  $\phi(x,\mu)$ satisfy the Brodsky-Lepage (BL)
evolution equation \cite{BroLep79,BroLep80,EfrRad80a}:
\begin{eqnarray}
\label{evo-equ-mes}
\mu^2 \frac{d}{d \mu^2} \phi(x,\mu) =
		 V(x,y;\alpha_s(\mu))\otimes\phi(y,\mu).
\end{eqnarray}
Of course, the physical quantities are independent of the chosen scheme,
which means that
\begin{eqnarray}
\label{RG-transfor-1}
T^{\rm CS}(\omega,x,Q,\mu)&=& T^{\rm MS}(\omega,y,Q,\mu) \otimes B(y,x,\mu),
\\
\phi^{\rm CS}(x,\mu)&=&B^{-1}(x,y,\mu)\otimes \phi^{\rm MS}(y,\mu),
\nonumber
\end{eqnarray}
where the $B$ kernels satisfy $B(x,z,\mu)\otimes B^{-1}(z,y,\mu)=\delta(x-y).$
The latter transformation implies that the evolution kernel of the DA
transforms inhomogeneously:    
\begin{eqnarray}
\label{RG-transfor-ker}
V^{\rm CS}(x,y)=B^{-1}(x,z)\otimes V^{\rm MS}(z,z')\otimes B(z',y) - 
	\left[\mu^2\frac{d}{d\mu^2}B^{-1}(x,z)\right]\otimes B(z,y).
\end{eqnarray}

Perturbative QCD predictions for physical quantities are given as truncated
series in $\alpha_s$. This necessary truncation induces their scheme
dependence. The unknown higher order corrections may be minimized by
choosing an appropriate scheme and scale. The problem of finding such an
optimal renormalization scheme can be attacked with the help of the extended
renormalization group equations introduced by St\"uckelberg and Peterman
\cite{LuBro93}, which is equivalent to previous work given in Refs.\
\cite{Gru80,Gru84,DhaGup84,GupShiTar91}. To our best knowledge, no
comparable methods were developed to find also the optimal factorization
scheme. For a given process there  may exist physical arguments to favour
a special scheme. A further requirement should be that the factorization
scheme respects the underlying symmetries of the theory, in order that no
anomalous terms appear either in the hard-scattering part or in the BL
evolution equation.

The more restricted problem to find the optimal scale in a given scheme has
been widely discussed in the literature and three quite distinct methods
have been proposed: the principle of fastest apparent convergence (FAC)
\cite{Gru80,Gru84}, the principle of minimal sensitivity (PMS)
\cite{Ste81,Ste81a,Ste82,Ste84}, and the Brodsky-Lepage-Mackenzie (BLM)
\cite{BroLepMac83} scale setting. The application of these methods can yield
quite different predictions (see for instance the analyses in Ref.\
\cite{KraLam91}).

Although conformal symmetry holds only true in the hypothetical conformal
limit, it should be manifested in the maximally possible manner in the full
theory. For exclusive processes in which only mesons participate such a
factorization scheme is, up to the scale setting problem, uniquely defined
in the conformal limit\footnote{In the case where baryons are also involved,
it is known that the mixing problem in the evolution equation cannot be
completely solved by conformal constraints \cite{Ohr82}, and thus
also in this conformal subtraction scheme some freedom remains.}. However,
as discussed above, also in the CS scheme anomalous terms proportional to
the $\beta$-function are left and cannot, as we will see below, be uniquely
fixed. According to the BLM scale setting prescription, these anomalous
terms can be absorbed in the scale setting of the coupling, and the
perturbative series will be formally the same as in the conformal theory.
Motivated by our discussion it seems worthwhile to reexamine the known NLO
corrections to exclusive processes in the CS scheme and to employ the BLM
prescription for the absorption of the remaining conformal anomalies.

\section{Evolution of the flavour non-singlet distribution amplitude}
\label{NLO-DA}
\subsection{General formalism}

First we discuss the application of the BLM scale fixing prescription in the
evolution of the DA. For the convenience of the reader we outline the whole
formalism for the solution of the evolution equation (\ref{evo-equ-mes}) in
terms of the conformal partial wave expansion. Using the
$\overline{\rm MS}$ scheme, the evolution kernel
\begin{eqnarray}
V(x,y;\alpha_s)=
	\frac{\alpha_s}{2\pi} V^{(0)}(x,y) +
	\left(\frac{\alpha_s}{2\pi}\right)^2 V^{(1)}(x,y) + \cdots
\end{eqnarray}
was computed perturbatively in one- and two-loop approximation
\cite{DitRad84,Sar84,MikRad85,Kat85}. The one-loop kernel is diagonal with
respect to Gegenbauer polynomials $C_k^{3\over2}(2x-1)$ of order $k$ and
with index $3/2$. In the two-loop approximation this property is spoiled and
because of the complicated structure, the moments
\begin{eqnarray}
\gamma_{kn}(\alpha_s)  = -2 \int_0^1 dx \int_0^1 dy\, C_k^{3\over2}(2x-1)
V(x,y;\alpha_s)  {(1-y)y \over N_{n}} C_{n}^{3\over 2}(2y-1),
\end{eqnarray}
where the normalization factor is $N_n={(n+1)(n+2)}/({4(2n+3)})$, cannot be
directly calculated. Fortunately, one can use conformal constraints to
compute the off-diagonal moments in a very economical way
\cite{Mue94,BelMue98a,BelMue98c}. Indeed, in the \mbox{ $\overline{\rm MS} $
} scheme the off-diagonal moments of the matrix $\hat{\gamma}$ are induced
by a special conformal anomaly matrix $\hat{\gamma}^c$ and the running
coupling \cite{Mue94,BelMue98a,BelMue98c}:
\begin{eqnarray}
\label{conf-constr-KD}
\left[\hat{a}(l)+\hat{\gamma}^c(l)
+2{\beta\over g}\hat{b}(l),\hat{\gamma}\right]=0,
\end{eqnarray}
where the matrices $\hat{a}$ and $\hat{b}$ have the following elements:
\begin{eqnarray}
\label{def-a}
a_{kn}(l)&= &2(k-l)(k+l+3) \delta_{kn},
\\
\label{def-b}
b_{kn}(l)&=&\left\{\begin{array}{c@{\quad}l} 
                          2(l+n+3)\delta_{kn} - 2(2n+3) & 
                          \mbox{if } k-n\ge 0 \mbox{ and even}\\
                             0     &   \mbox{otherwise.} 
                          \end{array}\right.
\end{eqnarray}
While the off-diagonal terms, induced by the renormalization of the
coupling, are predicted by conformal constraints (\ref{conf-constr-KD}), the
special conformal anomaly matrix $\hat{\gamma}^c$ contains new information.
As explained in Refs.\ \cite{Mue91a,Mue94,BelMue98a,BelMue98c}, this anomaly
matrix can be calculated directly with the help of modified Feynman rules
and it reads to leading order (LO):
\begin{eqnarray}
\label{CWI-scBre2}
\hat{\gamma}^{c(0)}(l)&=& -\hat{b}(l) \hat{\gamma}^{(0)}+\hat{w},
\nonumber\\
&&\hspace{-3.6cm}\mbox{\ where}
\nonumber\\
w_{kn}&=&C_F \left\{\begin{array}{c@{\quad}l} 
                          -4(2n+3)(k-n)(k+n+3)\times
                           & 
                            \mbox{if } k-n>0 \\
 \left[{A_{kn}-\psi(k+2)+\psi(1)\over (n+1)(n+2)} 
               +{2A_{kn}\over (k-n)(k+ n+3)}\right] 
                           & 
                           \mbox{and even}\\
                             0     &   \mbox{otherwise} 
                          \end{array}\right. ,
\\
A_{kn}&=&
		 \psi\left({k+n+4\over 2}\right)-\psi\left({k-n\over 2}\right)
       +2\psi\left(k-n\right)-\psi\left(k+2\right)-\psi(1), 
\nonumber
\end{eqnarray}
with $\psi\left(z\right)= {d\over dz} \ln\Gamma(z)$ and $C_F=4/3$.

Employing the conformal partial wave expansion, which is given in terms
of the eigenfunctions of the LO kernel $ V^{(0)}(x,y)$,
\begin{eqnarray}
\label{DA-ConSpin-Exp-4}
\phi(x,Q) =
	\sum_{k=0}^{\infty}{^\prime} {(1-x)x \over N_k} C_k^{3\over 2}(2x-1)  
	\langle 0| O_{kk}(\mu)|M(P) \rangle^{red}_{|\mu=Q},
\end{eqnarray}
the BL evolution equation (\ref{evo-equ-mes}) can be perturbatively solved
to any order \cite{Mue94}. Note that the expectation values of the operators
for odd $k$ vanish, which is indicated by the $\sum{^\prime}$ symbol. The
evolution of the composite operators appearing in Eq.\
(\ref{DA-ConSpin-Exp-4}) is governed by the renormalization group equation
(RGE), which possesses the following triangular form due to Poincar\'e
invariance:
\begin{eqnarray}
\label{RGE-CS-5}
\mu\frac{d}{d\mu} O_{kl}=
				-\gamma_k(\alpha_s(\mu))O_{kl}
				-\sum_{n=0}^{k-2}{^\prime}\gamma_{kn}^{\rm ND}(\alpha_s(\mu))
				O_{nl}.
\end{eqnarray}
The off-diagonal matrix elements $\gamma^{\rm ND}_{kn}$ (with $k>n$) appear
beyond the LO and are scheme dependent. In the $\overline{\rm MS} $ scheme
they can be simple obtained from Eqs.\
(\ref{conf-constr-KD})--(\ref{CWI-scBre2}).

In Ref.\ \cite{Mue97a} it has been shown that in the conformal limit of the
theory the CS scheme ensures the conformal covariance of the renormalized
operators and, therefore, their anomalous-dimension matrix is diagonal. The
transformation to the CS scheme is determined by a matrix $\hat{B}$, which
depends only on $\hat{\gamma}^c$:
\begin{eqnarray}
\label{blefdt-1}
\hat{B} = {\hat{1} \over \hat{1}+{\cal J}\hat{\gamma}^c}
   = \hat{1}-{\cal J}\hat{\gamma}^c + {\cal J}(\hat{\gamma}^c 
            {\cal J}\hat{\gamma}^c)-\cdots,
\end{eqnarray}
where the operator ${\cal J}$ is defined by
\begin{eqnarray}
{\cal J}\hat{A}:=
\left\{\begin{array}{c@{\quad}l} 
                  {A_{kn}\over 2(k-n)(k+n+3)}        
                           & 
                            \mbox{if } k-n >0 \\
						0     &   \mbox{otherwise.} 
                          \end{array}\right.
\end{eqnarray}
This transformation
cancels the off-diagonal part of $\hat\gamma^{\rm MS}$, however, it induces
an off-diagonal term proportional to the $\beta$-function:
\begin{eqnarray}
\label{AnoDim-CS-MS}
\hat{\gamma}^{\rm CS}= 
	\hat{B}^{-1}\hat{\gamma}^{\rm MS} \hat{B} - \beta
	\left[\frac{\partial}{\partial g}\hat{B}^{-1}\right]\hat{B}.
\end{eqnarray}
Hence, in this
scheme the off-diagonal part of the anomalous-dimension matrix is
proportional to the $\beta$-function:
\begin{eqnarray}
\label{RGE-CS-offdiagonal}
\gamma_{kn}^{\rm ND} = \frac{\beta}{g}\Delta_{kn},\quad
\mbox{where\ } \Delta_{kn} =
\frac{\alpha_s}{2\pi} \Delta^{(0)}_{kn} + O(\alpha_s^2).
\end{eqnarray}

The BLM scale setting prescription \cite{BroLepMac83} can now be applied
to absorb the off-diagonal term (\ref{RGE-CS-offdiagonal}) into the scale
dependence of the coupling:
\begin{eqnarray}
\label{alphas-ScaRel}
\alpha_s(\mu^\ast) = \alpha_s(\mu)\left[1-
\beta_0 \frac{\alpha_s(\mu)}{2\pi}
\ln\left(\frac{\mu^\ast}{\mu}\right) + \cdots \right].
\end{eqnarray}
Thus, the off-diagonal term in NLO may be expressed as
\begin{eqnarray}
\label{RGE-CS-BLM-verification}
\gamma_{kn}^{\rm ND} &=&
-\left(\frac{\alpha_s(\mu)}{2\pi}\right)^2 \beta_0\Delta^{(0)}_{kn} + \cdots
=
\left[\frac{\alpha_s(\mu^\ast_{kn})}{2\pi}-\frac{\alpha_s(\mu)}{2\pi}\right]
\gamma^{(0)}_k+ \cdots
\\
&=&
\gamma_k(\alpha_s(\mu^\ast_{kn}))-\gamma_k(\alpha_s(\mu)) + \cdots,
\nonumber
\end{eqnarray}
where 
$\mu^\ast_{kn}=\mu \exp\left\{\Delta^{(0)}_{kn}/\gamma^{(0)}_k\right\}$
is the new scale. To absorb the NNLO corrections in the analogous way, it is
necessary to introduce a different second scale $\mu^{\ast\ast}_{kn}$
and so on. After this procedure the RGE (\ref{RGE-CS-BLM}) in the
CS scheme takes the form:
\begin{eqnarray}
\label{RGE-CS-BLM}
\mu\frac{d}{d\mu} O^{\rm co}_{kl}=
						-\gamma_k(\alpha_s(\mu))O^{\rm co}_{kl}
						-\sum_{n=0}^{k-2}{^\prime}\left[
\gamma_k(\alpha_s(\mu^\ast_{kn}),\alpha_s(\mu^{\ast\ast}_{kn}),\dots)-
						\gamma_k(\alpha_s(\mu))
											\right]	O^{\rm co}_{nl}.
\end{eqnarray} 
All scales $\mu^\ast_{kn}, \mu^{\ast\ast}_{kn},\dots$ are uniquely
determined by the conformal symmetry breaking term $\Delta_{kn}$. This
anomaly arises from two sources: (i) from the renormalization of the
coupling constant that enters in the anomalous-dimension matrix and (ii)
from the renormalization prescription in the CS scheme.

Here one comment is in order. Since the effects of arbitrary renormalization
transformations proportional to the $\beta$ function will disappear in the
conformal limit, the CS scheme cannot be fixed in the full theory. Below we
will deal with two such CS schemes referred as $\mbox{CS}_{\rm I}$ and
$\mbox{CS}_{\rm II}$. In the first one only the off-diagonal part that is
related to the special-conformal anomaly matrix will be removed by the
transformation (\ref{AnoDim-CS-MS}), while in the $\mbox{CS}_{\rm II}$
scheme also the off-diagonal terms proportional to $\beta_0$ appearing in
the anomalous dimensions are explicitly removed (obviously, the
inhomogeneous part in Eq.\ (\ref{AnoDim-CS-MS}) cannot be avoided, and
finally terms proportional to $\beta_0$ appear). Although the
$\mbox{CS}_{\rm II}$ scheme induces an additional symmetry breaking term
proportional to the $\beta$-function in the hard-scattering part, which
looks artificial, it will be interesting to compare both CS schemes with
each other.

The evolution equation (\ref{evo-equ-mes}) can be perturbatively solved with
the help of the conformal spin expansion. Solving the RGE
(\ref{RGE-CS-5}), which is an inhomogeneous partial first order differential
equation, the solution has been written in a compact form in Ref.\
\cite{Mue94} and it is valid for an arbitrary scheme:
\begin{eqnarray}
\label{DA-ConSpin-Exp-CS-fin}
\phi(x,Q^2) =
	\sum_{n=0}^{\infty}{^\prime}\varphi_n(x,Q,Q_0)
	\exp\left\{-\int_{Q_0}^Q\frac{d\mu}{\mu}\gamma_{n}(\mu)\right\}
	\langle 0| O_{nn}(Q_0)|M(P) \rangle^{red}.
\end{eqnarray} 
The partial waves $\varphi_n(x,Q,Q_0)$ now contain  perturbative corrections,
which are induced by the off-diagonal anomalous dimension matrix. They are
also known as expansion with respect to the Gegenbauer polynomials
\begin{eqnarray}
\label{DA-ConSpin-Exp-CS-PW}
\varphi_n(x,Q,Q_0)=
		\sum_{k=n}^{\infty}{^\prime} 
		{(1-x)x \over N_k} C_k^{3\over 2}(2x-1)  B_{kn}^{\rm dia}(Q,Q_0). 
\end{eqnarray}
The matrix $B_{kn}^{\rm dia}(Q,Q_0)$ diagonalizes the RGE (\ref{RGE-CS-5})
of the conformal operators and is given by
\begin{eqnarray}
\label{sol-B-mixing}
\hat{B}^{\rm dia}=
	\frac{\hat{1}}{\hat{1}-{\cal L}\hat{\gamma}^{\rm ND}}
	=\hat{1}+{\cal L}\hat{\gamma}^{\rm ND}+
	 {\cal L}\left(\hat{\gamma}^{\rm ND}{\cal L}\hat{\gamma}^{\rm
	ND}\right)+\cdots,
\end{eqnarray}
where the operator ${\cal L}$ is an integral operator
acting on a triangular and off-diagonal matrix:
\begin{eqnarray}
\label{sol-B-mixing-L}
{\cal L}\gamma^{\rm ND}_{kn}=
	-\int_{Q_0}^Q \frac{d\mu}{\mu} \gamma^{\rm ND}_{kn}(\mu)
\exp\left\{
-\int_{\mu}^Q\frac{d\mu'}{\mu'}\left[\gamma_{k}(\mu')-
	\gamma_{n}(\mu')\right] \right\}.
\end{eqnarray}
Here we do not include radiative corrections at the reference point $Q_0$,
i.e. $B^{\rm dia}_{kn}(Q_0,Q_0)=\delta_{kn}$. Notice that in general the
matrix $\hat{B}^{\rm dia}$ is different from the previously introduced
matrix $\hat{B}$.

\subsection{NLO analysis}
\label{NLO-DA-B}

In the $\overline{\mbox{MS}}$ scheme the evolution of the DA in NLO was
analysed in detail in Ref.\ \cite{Mue95}. The main feature showing up in this
order is the excitation of higher harmonics due to the mixing of the
operators, which yields logarithmic corrections in the end-point region.
This logarithmical enhancement is hidden in the expansion of the partial
waves, which is, corresponding to Eqs.\
(\ref{sol-B-mixing})-(\ref{DA-ConSpin-Exp-CS-PW}), given as
\begin{eqnarray}
\label{DA-Evo-MS}
\varphi_n(x,Q,Q_0)&=&
	{(1-x)x \over N_n} C_n^{3\over 2}(2x-1)+
	\frac{\alpha_s(Q)}{2\pi}\varphi_n^{(1)}(x,Q,Q_0)+\cdots,
\\
\varphi_n^{(1)}(x,Q,Q_0)&=&
		\sum_{k=n+2}^{\infty}{^\prime} 
		{(1-x)x \over N_k} C_k^{3\over 2}(2x-1)
		\frac{1-\left(\frac{\alpha_s(Q_0)}{\alpha_s(Q)}\right)^{
		\frac{\beta_0+\gamma_n^{(0)}-\gamma_k^{(0)}}{\beta_0}}}
			 {\beta_0+\gamma_n^{(0)}-\gamma_k^{(0)}}
    \gamma_{kn}^{(1){\rm ND}}.
\nonumber
\end{eqnarray}
The off-diagonal matrix $\hat{\gamma}^{(1){\rm ND}}$ is determined by
the conformal anomalies and  in the $\overline{\mbox{MS}}$ scheme it can be
easily obtained from  Eq.\ (\ref{conf-constr-KD}):
\begin{eqnarray}
\gamma_{kn}^{(1){\rm ND}}=
	(\gamma_k^{(0)}-\gamma_n^{(0)})\frac{\gamma^{c(0)}_{kn}-\beta_0 b_{kn}}
					{2(k-n)(k+n+3)}.
\end{eqnarray}
Here the matrices $\hat{b}$ and $\hat{\gamma}^{c}$ are defined in
Eqs.\ (\ref{def-b}) and (\ref{CWI-scBre2}), respectively.

Above we introduced two conformal schemes $\mbox{CS}_{\rm I}$ and
$\mbox{CS}_{\rm II}$ which are obtained from the $\overline{\rm MS} $ one by
the transformations
\begin{eqnarray}
\label{tranf-MS-CS}
B_{kn}&=&\delta_{kn}-
	\frac{\alpha_s}{2\pi}\frac{\gamma^{c(0)}_{kn}}{2(k-n)(k+n+3)}\quad
\mbox{for CS}_{\rm I},
\\
B_{kn}&=&\delta_{kn}-
	\frac{\alpha_s}{2\pi}\frac{\gamma^{c(0)}_{kn}-
								\beta_0 b_{kn}}{2(k-n)(k+n+3)}\quad
\mbox{for CS}_{\rm II},
\nonumber
\end{eqnarray}
respectively. According to Eq.\ (\ref{AnoDim-CS-MS}) these transformations
imply that the off-diagonal part is $\gamma_{kn}^{(1){\rm ND}}= -\beta_0
\Delta_{kn}^{(0)}$ [see Eq.\ (\ref{RGE-CS-offdiagonal})] with
\begin{eqnarray}
\Delta_{kn}^{(0)}=
\frac{(\gamma_k^{(0)}-\gamma_n^{(0)})b_{kn}-\gamma^{c(0)}_{kn}}{2(k-n)(k+n+3)}
					\quad\mbox{for CS}_{\rm I},
\\
\Delta_{kn}^{(0)}=
\frac{\beta_0 b_{kn} -\gamma^{c(0)}_{kn}}{2(k-n)(k+n+3)}
					\quad\mbox{for CS}_{\rm II}.
\nonumber
\end{eqnarray}
$\Delta_{kn}^{(0)}$ can be absorbed via the BLM scale
setting prescription; so that then the partial waves read
\begin{eqnarray}
\varphi^{\rm CS}_n(x,Q,Q_0)&=&
	{(1-x)x \over N_n} C_n^{3\over 2}(2x-1)+
 		\sum_{k=n+2}^{\infty}{^\prime} 
		{(1-x)x \over N_k} C_k^{3\over 2}(2x-1)\gamma_k^{(0)}
		\left(1-\frac{\alpha_s(Q)}{\alpha_s(Q^\ast_{kn})}\right)
\nonumber\\
&&\hspace{4,5cm} \times 
		\frac{1-\left(\frac{\alpha_s(Q_0)}{\alpha_s(Q)}\right)^{
		\frac{\beta_0+\gamma_n^{(0)}-\gamma_k^{(0)}}{\beta_0}}}
			 {\beta_0+\gamma_n^{(0)}-\gamma_k^{(0)}}+\cdots. 
\end{eqnarray}

Finally, we study in the $\overline{\rm MS}$ and in both CS schemes the
scale dependence of the quantity
\begin{eqnarray}
\label{I-integral}
I(Q,Q_0) = \int_0^1 dx\, \frac{\varphi(x,Q,Q_0)}{x},
\end{eqnarray}
which enters in different exclusive mesonic processes. For convenience we
changed the normalization of the DA, i.e. $\phi(x)= f_\pi
\varphi(x)/2\sqrt{6}$, where the pion decay constant is $f_\pi\approx 0.131$
GeV. Notice that this integral was originally defined to LO; here, however,
it also contains the higher order effects caused by the evolution. It is
very sensitive to the end-point behaviour of the DA and, therefore, it may
serve as a measure for the logarithmic corrections due to the evolution,
which occur in this region \cite{Mue95}.

Inserting the conformal spin expansion (\ref{DA-ConSpin-Exp-CS-fin}) into the
integral (\ref{I-integral}) provides the representation
\begin{eqnarray}
\label{1ovex-Evo}
I(Q,Q_0) &=&
	\sum_{n=0}^{\infty}{^\prime}	I_n(Q,Q_0)
	\exp\left\{-\int_{Q_0}^Q\frac{d\mu}{\mu}\gamma_{n}(\mu)\right\}
	\langle 0| O_{nn}(Q_0)|\pi(P)\rangle^{red},
\\
\label{1ovex-Evo-PW}
I_n(Q,Q_0)&=&\frac{2(3+2n)}{(n+1)(n+2)}+
	 			\frac{\alpha_s(Q)}{2\pi} \sum_{k=n+2}^{\infty}{^\prime} 
		\frac{2(3+2k)}{(k+1)(k+2)}\gamma_{kn}^{(1){\rm ND}}
\\
&&\hspace{5,5cm} \times 
		\frac{1-\left(\frac{\alpha_s(Q_0)}{\alpha_s(Q)}\right)^{
		\frac{\beta_0+\gamma_n^{(0)}-\gamma_k^{(0)}}{\beta_0}}}
			 {\beta_0+\gamma_n^{(0)}-\gamma_k^{(0)}}+\cdots,
\nonumber
\end{eqnarray}
which will be evaluated numerically by an appropriate truncation of the
series. Here the reduced matrix elements are the conformal moments of the DA
now normalized as: 
\begin{eqnarray}
\label{new-normal}
\langle 0| O_{nn}(\mu)|\pi(P)\rangle^{red}=
 \int_0^1dx\, C_n^{\frac{3}{2}}(2x-1) \varphi(x,\mu).
\end{eqnarray}
It is worthy to mention that in the CS schemes, the direct
integration of the RG equation (\ref{RGE-CS-BLM}) implies the following
form:
\begin{eqnarray}
\label{1ovex-Evo-PW-CS}
I_n(Q,Q_0)&=&\frac{2(3+2n)}{(n+1)(n+2)}
			\left(1+1-\frac{\alpha_s(Q)}{\alpha_s(\overline{Q}_{n})}+
			\cdots\right),
\end{eqnarray}
where $\overline{Q}_{n}$ depends on $Q$ and $Q_0$:
\begin{eqnarray}
1-\frac{\alpha_s(Q)}{\alpha_s(\overline{Q}_{n})}&=& 
				\frac{(n+1)(n+2)}{2(3+2n)} \sum_{k=n+2}^{\infty}{^\prime} 
				\frac{2(3+2k)}{(k+1)(k+2)} \gamma_k^{(0)}
		\left(1-\frac{\alpha_s(Q)}{\alpha_s(Q^\ast_{kn})}\right)
\\
&&\hspace{6,5cm} \times 
		\frac{1-\left(\frac{\alpha_s(Q_0)}{\alpha_s(Q)}\right)^{
		\frac{\beta_0+\gamma_n^{(0)}-\gamma_k^{(0)}}{\beta_0}}}
			 {\beta_0+\gamma_n^{(0)}-\gamma_k^{(0)}}.
\nonumber
\end{eqnarray}
Notice that in the conformal scheme $\mbox{CS}_{\rm II}$ a contribution of
the form $1-\alpha_s(Q)/\alpha_s(\overline{Q})$ arises in the
hard-scattering amplitudes, too. Obviously, their expansion gives an
$\alpha_s$-suppressed term proportional to $\beta_0$.

Taking into account a sufficient large number of terms in the series
(\ref{1ovex-Evo}) -- (\ref{1ovex-Evo-PW-CS}) (details for the estimate of
the accuracy of such an approximation can be found in Ref.\ \cite{Mue95}), the
relative correction to NLO, i.e.
\begin{eqnarray}
I^{\rm rel}(Q,Q_0)=
	\frac{I^{\rm NLO}(Q,Q_0)-I^{\rm LO}(Q,Q_0)}
	{I^{\rm LO}(Q,Q_0)},
\end{eqnarray}
is numerically evaluated, where we consequently took into account the
following perturbative expansion:
\begin{eqnarray}
\exp\left\{-\int_{Q_0}^Q\frac{d\mu}{\mu}\gamma_{n}(\mu)\right\} &=&
\left(\frac{\alpha_s(Q)}{\alpha_s(Q_0)}\right)^\frac{\gamma_n^{(0)}}{\beta_0}
\left[1+\left(\frac{\alpha_s(Q)}{2\pi} -\frac{\alpha_s(Q_0)}{2\pi}\right)
\left(\frac{\gamma_n^{(1)}}{\beta_0}-
\frac{\beta_1}{2\beta_0}\frac{\gamma_n^{(0)}}{\beta_0}\right)+\cdots \right].
\nonumber\\
&&\hspace{-15cm}{}
\end{eqnarray}
Here $\beta_0=11-2/3 n_f$, $\beta_1=102-38/3 n_f$, $\gamma_n^{(0)}= -C_F
\{3+1/[(n+1)(n+2)]-4\psi(n+2)+4\psi(1)\}$, and the expression for
$\gamma_n^{(1)}$ can be found for instance in Ref.\ \cite{GonLopYnd79}. Two
values of the QCD scale parameter, a rather small one $\Lambda^{\rm LO}=100$
MeV as well as a rather large one $\Lambda^{\rm LO}=500$ MeV are chosen.
Since in general non-perturbative (model) calculations are performed at a
low scale, we choose the reference momentum $Q_0=\sqrt{0.5}\mbox{\ GeV}$. In
the following it is assumed that the evolution equation can be used at this
low scale. The evolution runs up to the scale $Q=\sqrt{20} \mbox{\ GeV}$.
The number of active quarks is $n_f=3$ and we take the following set of DAs:
asymptotic, two-hump and $\varphi^{a}(x)=\Gamma(2a+2)/\Gamma(a+1)^2
[x(1-x)]^a$ with $a=\{10,1/2,1/4\}$, i.e. very narrow and broad ones are
also included. The resulting relative NLO corrections in percentage are
listed for the three considered schemes in Table \ref{tab-I-evo}.


\begin{table}[htb]
\begin{center}
\begin{tabular}{|c||c||r@{.}l|r@{.}l|r@{.}l|r@{.}l|r@{.}l|r@{.}l|}\hline
\multicolumn{1}{|c||}{\ }&\multicolumn{1}{c||}{\ }
 	&\multicolumn{6}{c|}{$\Lambda=100$ MeV}
	&\multicolumn{6}{c|}{$\Lambda=500$ MeV}\\
\multicolumn{1}{|c||}{DA}&\multicolumn{1}{c||}{$I(Q_0,Q_0)$}
&\multicolumn{2}{c|}{$\overline{\rm MS}$}
&\multicolumn{2}{c|}{$\mbox{CS}_{\rm I}$}
&\multicolumn{2}{c|}{$\mbox{CS}_{\rm II}$}
&\multicolumn{2}{c|}{$\overline{\rm MS}$}
&\multicolumn{2}{c|}{$\mbox{CS}_{\rm I}$}
&\multicolumn{2}{c|}{$\mbox{CS}_{\rm II}$}\\
\hline\hline
$\varphi^{10}$ 	&2.1& 0&7\% &1&0\% &1&0\%	&1&6\% 	&4&2\%  &2&2\%\\\hline
$\varphi^{\rm as}$	&3.0&-0&1\% &0&4\% &-0&4\% 	&-1&3\% &2&3\%
															&-3&0\%\\\hline
$\varphi^{1/2} $ 	&4.0&-0&8\%	&-0&3\%	&-2&4\%  &-3&1\% &0&7\%
															&-7&8\%\\\hline
$\varphi^{\rm CZ}$	&5.0&-1&2\%	&-0&4\%	&-2&5\% &-5&6\%  &-0&6\%
															&-11&7\%\\\hline
$\varphi^{1/4} $ 	&6.0&-1&2\%	&-2&3\%	&-6&6\%  &-6&7\% &-1&6\%
														 &-25&\%\\\hline
\end{tabular}
\end{center}
\caption[Relative NLO corrections to the integral  $I$.]{
The value of the $I$ integral at the input scale $Q_0=\sqrt{0.5}$ GeV and
its relative NLO corrections at $Q=\sqrt{20}$ GeV are listed for different
DAs in the $\overline{\rm MS} $, ${\rm CS}_{\rm I}$, and $\mbox{CS}_{\rm II}
$ scheme for two choices of the QCD scale parameter $\Lambda^{\rm
LO}={100,{\ }500}$ MeV.}
\label{tab-I-evo}
\end{table}

For both the narrow and the asymptotic DA the NLO corrections are rather
small in all three schemes. The corrections are negative for broader DAs and
their absolute value is increasing with growing value of $I^{\rm LO}$,
however, they remain rather small in the $\mbox{CS}_{\rm I} $ scheme. In the
$\mbox{CS}_{\rm II} $ scheme the corrections are larger than in the
$\overline{\rm MS} $ scheme, so one may conclude that this conformal scheme
is disfavoured. Let us mention that in the $\overline{\mbox{MS}}$ scheme the
contributions from the off-diagonal part are of the same sign as the
$\alpha_s^2$ corrections to the diagonal part of the anomalous-dimension
matrix and that both of them enter in the relative NLO contribution with a
similar size. In the $\overline{\rm MS} $ and the $\mbox{CS}_{\rm I} $
scheme, but not in the $\mbox{CS}_{\rm II} $ one, a partial cancellation in
the off-diagonal terms takes place between the special conformal anomaly and
the $\beta_0$ term \cite{Mue95}, and that the net contribution is small.
Note that for a reference momentum of $1$ GeV or even larger the size of the
NLO corrections decreases. We may conclude that in general the NLO
corrections due to the evolution of the DA are small, except for very broad
DA evolved in the $\mbox{CS}_{\rm II} $ scheme.

\section{Photon-to-pion transition form factor}
\label{TFF}

From the theoretical point of view the simplest mesonic process is the
production of a pseudoscalar meson in two-photon collisions
\begin{eqnarray}
\gamma^{\ast}(q_1)\; \gamma^\ast(q_2)  \to M(P),
\end{eqnarray}
since it is purely electromagnetic to LO. As independent kinematical
variables we choose the negative of the momentum transfer between the
photons squared $Q^2=-q^2$, with $q=(q_1-q_2)/2$ and the asymmetry parameter
$\omega= Pq/Q^2$. In the case that one photon is on mass-shell we have
$|\omega|=1$, while for equal photon virtualities $|\omega|=0$. The
dynamical information is contained in the amplitude $\Gamma_{\alpha\beta}=
-i e^2\epsilon_{\alpha\beta\mu\nu} q_1^\mu q_2^\nu F_{\gamma
M}(\omega,Q^2)$, where the photon-to-meson transition form factor $F_{\gamma
M}(\omega,Q^2)$ is defined in terms of the time ordered product of two
electromagnetic currents sandwiched between the one-meson state and the
vacuum:
\begin{eqnarray}
\label{TFF-Def-ForFac}
-i e^2\epsilon_{\alpha\beta\mu\nu} q_1^\mu q_2^\nu F_{\gamma M}(\omega,Q^2)
=
	i \int d^4 x\, e^{ixq}
\bigg\langle M\bigg|T
	J^\mu\left(\frac{x}{2}\right) J^\nu\left(-\frac{x}{2}\right)
\bigg|0\bigg\rangle.
\end{eqnarray}

At large momentum transfer this transition form factor has been measured for
$\pi^0$, $\eta$, and $\eta^\prime$ mesons in single antitagged experiments
by the CELLO collaboration \cite{CELLO91} and, more recently, at CLEO
\cite{Sav95,Sav97}, where the untagged photon is almost real. The
photon-to-pion transition form factor has been determined for $0.5\le Q^2\le
2.7\mbox{\ GeV}^2$\ \cite{CELLO91} and $1.5\le Q^2\le 9\mbox{\ GeV}^2$,
respectively \cite{Sav95,Sav97}. For the CLEO data the virtuality of the
second photon was estimated to be less than $0.001\ \mbox{GeV}^2$. The
$\eta$ and $\eta^\prime$ transition form factors are known up to
$20$\mbox{$\mbox{\ GeV}^2$}\ and $30$\mbox{$\mbox{\ GeV}^2$} , respectively
\cite{Sav95,Sav97} (the systematic errors become very large with increasing
photon virtuality). Data at lower momentum transfer are given in Refs.\
\cite{Pluto84,TPC90}.

Taking the SFA to LO \cite{BroLep80,BroLep81}, the normalization of the
photon-to-pion transition form factor is consistent with pion DA's that are
{\em not} concentrated in the end-point region. Perturbative and
non-perturbative corrections to this prediction have been studied in a
number of papers. Let us only mention that non-perturbative effects have
been included in a model dependent way by the transverse momentum dependence
\cite{JakKroRau96,KroRau96,Ong95,CaoHuaMa96} or by the sum rule approach
\cite{RadRus96,RadRus96a,MusRad97}. All these analyses show that the data
can be reproduced by a DA that is close to the asymptotic one\footnote{In
the prediction of the transition form factor only the integral $\int_0^1
dx\, \varphi(x)/x$ enters and its value extracted from the data is of about
3 or even smaller. Of course, the asymptotic DA provides 3, however, we
should mention that one cannot conclude in the mathematical sense that its
shape is convex; even if one takes into account normalization and
positivity.}: $\varphi^{\rm as} =6x(1-x)$. It is maybe interesting to note
that the authors in Ref.\ \cite{CaoHuaMa96} observed that, taking into
account the transverse momentum dependence in the light-cone formalism,
quite different model wave functions are consistent with the data. In the
following we rely on the SFA, which allows us to calculate the perturbative
corrections in a systematic way \cite{BroLep80}. From this factorization
procedure it is expected that large soft corrections from the transverse
momentum dependence in the pion wave function would induce large
perturbative corrections in the SFA.

Alternative to the SFA the transition form factor at large momentum
transfer can be calculated with the help of the operator product expansion
(OPE) at light-like distances. In a conformal invariant theory, the form of
the Wilson coefficients are fixed up to the normalization
\cite{FerGriGat71a,FerGriGat72a}. In Ref.\ \cite{Mue97a} it has been shown
that this conformally covariant OPE, holds true for $\beta=0$ in the CS
scheme. Taking into account the general structure of this conformal
prediction, the decomposition of the transition form factor in conformal
partial waves is given in the conformal limit as:
\begin{eqnarray}
\label{TFF-Com-cOPE-fin}
F_{\gamma\pi}(\omega,Q)&=&
		\frac{2\sqrt{2}f_\pi}{3 Q^2}\sum_{k=0}^\infty {^\prime}
	B(k+1,k+2)c_k(\alpha_s(\mu))  \left(
				\frac{\mu^2}{(1+\omega)Q^2}
				\right)^{\frac{\gamma_k}{2}}
	\frac{2(2\omega)^{k}}{(1+\omega)^{k+1}}
\\    
&&\hspace{1cm} \times
	{_{2}F}_1\left({k+1+\frac{1}{2}\gamma_k, k+2+\frac{1}{2}\gamma_k
 \atop
2(k+2+\frac{1}{2}\gamma_k)}\Bigg|\frac{2\omega}{1+\omega}\right)
		\langle \pi(P)| O^{\rm co}_{kk}(\mu)|0\rangle^{red}.
\nonumber
\end{eqnarray}
The Wilson coefficients $c_k$ and the anomalous dimensions $\gamma_k$
are known up to order $\alpha_s^2$ from the perturbative corrections
to the longitudinal structure function $g_1$ \cite{ZijNee94,CurFurPet80}.
Here the reduced matrix elements are normalized according to Eq.\
(\ref{new-normal}) and  $c_k$ are equal to one in LO.

As already mentioned, the pion transition form factor has been measured by
single antitagged experiments in the region of $0.5-9 \mbox{\ GeV}^2$ at
CELLO and CLEO, where the antitagged photon is almost on-shell, so that
we can set $\omega=1$. Hence, employing the relation
${_{2}F}_1(a, b, c|1)=
  \frac{\Gamma(c)\Gamma(c-a-b)}{\Gamma(c-a)\Gamma(c-b)}$
leads to a considerable simplification of the prediction:  
\begin{eqnarray}
\label{TFF-Com-cOPE-fin-onshell}
F_{\gamma\pi}(\omega=1,Q^2)&=&
\frac{2 \sqrt{2} f_\pi}{3 Q^2}\sum_{k=0}^\infty {^\prime} \frac{
	\Gamma(k+1)\Gamma(k+2)\Gamma\left(2k+4+\gamma_k\right)}
	{\Gamma\left(k+2+\frac{1}{2}\gamma_k\right)
	 \Gamma\left(k+3+\frac{1}{2}\gamma_k\right)\Gamma\left(2k+3\right)}
\\
&&\hspace{3cm}\times c_k(\alpha_s(\mu))
	\left(\frac{\mu^2}{ -q_1^2}\right)^{\frac{\gamma_k}{2}} 
	\langle \pi(P)| O^{\rm co}_{kk}(\mu^2)|0\rangle^{red}.
\nonumber 
\end{eqnarray}

\subsection{NLO corrections}

It is again worthwhile to mention that in the favoured \mbox{$\mbox{CS}_{\rm
I} $ } scheme the prediction (\ref{TFF-Com-cOPE-fin-onshell}) is exact to
NLO, while in the \mbox{ $\overline{\rm MS} $ } scheme and the
\mbox{$\mbox{CS}_{\rm II} $ } scheme it is spoiled by the special conformal
anomaly and a term proportional to $\beta_0$, respectively. To evaluate the
formula (\ref{TFF-Com-cOPE-fin-onshell}) to NLO in the \mbox{$\mbox{CS}_{\rm
I} $ } scheme, the right-hand side (RHS) of Eq.\
(\ref{TFF-Com-cOPE-fin-onshell}) is consequently expanded up to the order
$\alpha_s$. The NLO results in the two other schemes are easily obtained by
employing the transformations (\ref{tranf-MS-CS}). Table \ref{tab-TFF-HS}
contains the absolute and relative $\alpha_s$ corrections to the
hard-scattering amplitude, where the virtuality of the space-like photon is
$-q^2_1=2 \mbox{\ GeV}^2$ and the QCD scale parameter is set to
$\Lambda^{\rm LO}=0.22$ GeV. For simplicity, we use the ``natural'' scale
setting prescription $\mu^2=-q_1^2$ in Eq.\ (\ref{TFF-Com-cOPE-fin-onshell})
[see the $\mu$ dependence of the coefficient function] and do not discuss
the scale setting dependence. The set of DAs is the same as before, however,
now at the input scale of $\sqrt{2}$ GeV instead of 0.5 GeV, with the
exception that the two-hump function is evolved from its normalization point
at 0.5 GeV to $\sqrt{2}$ GeV \cite{CheZhi82}. For the narrow DA the NLO
correction is in all schemes surprisingly large, namely, between $-25$ to
$-31\%$. For the asymptotic DA the NLO correction is about $-20\%$ in both
the \mbox{ $\overline{\rm MS} $ } and the \mbox{$\mbox{CS}_{\rm II} $ }
scheme, while it is only $-12\%$ in the \mbox{$\mbox{CS}_{\rm I} $ } scheme.
A similar reduction of about $10\%$ in the favoured \mbox{$\mbox{CS}_{\rm I}
$ } scheme is observed in the case of the two-hump DA resulting in
negligible small perturbative corrections in this case. For convex
amplitudes that are strongly concentrated in the end-point region the
corrections are positive and becoming very large. These results are due to
the fact that {\em only} the {\em first} conformal partial wave has a
negative contribution.

\begin{table}[htb]
\begin{center}
\begin{tabular}{|c||c||c|c|c||r@{\%}|r@{\%}|r@{\%}|}\hline
DA  &\multicolumn{1}{c||}{\mbox{LO}}&\multicolumn{6}{c|}{NLO}\\
\mbox{\ }  &\multicolumn{1}{c||}{\mbox{\ }}&\multicolumn{1}{c|}
{$\overline{\rm MS}$}&\multicolumn{1}{c|}
{$\mbox{CS}_{\rm I}$}&\multicolumn{1}{c||}{$\mbox{CS}_{\rm II}$}
&\multicolumn{1}{c|}{$\overline{\rm MS}$}&\multicolumn{1}{c|}
{$\mbox{CS}_{\rm I}$}&\multicolumn{1}{c|}{$\mbox{CS}_{\rm II}$}\\
\hline\hline
$\varphi^{10}$ 	&0.13&0.09&0.10&0.09	& -31 &-25  & -28	 \\\hline
$\varphi^{\rm as}$	&0.19&0.15&0.16&0.14	&-20  &-12  &-21      \\\hline
$\varphi^{1/2} $ 	&0.25&0.26&0.27&0.24	&5    &11   &-5        \\\hline
$\varphi^{\rm CZ}$	&0.26&0.24&0.26&0.22	&-10  & -1	&-14     \\\hline
$\varphi^{1/4} $ 	&0.37&0.78&0.76&0.63	&110  &104	&70       \\\hline
\end{tabular}
\end{center}
\caption[Predictions for the photon-to-pion transition form factor
		to LO and NLO.]{
Absolute LO and NLO predictions for the photon-to-pion transition form
factor and their relative deviation are listed for different DAs. Here the
scale $Q$ is set to $\sqrt{2}$ GeV and the evolution of the DA is 
taken into account only for the two-hump function.
}
\label{tab-TFF-HS}
\end{table}

Here one comment is in order. For a {\em given} DA, the differences of the
predictions in Table \ref{tab-TFF-HS} do {\em not} reflect the scheme
dependence of the NLO result. One has also to take into account
the transformation (\ref{RG-transfor}) of the DA, which would reduce the
observed differences drastically. For instance, if we take the asymptotic DA
in the \mbox{$\mbox{CS}_{\rm I} $ } scheme, the input in the
$\overline{\mbox{MS}}$ scheme is \cite{Mue95}
\begin{eqnarray}
\varphi^{\rm MS}(x)= B\otimes 6x(1-x)
		=6x(1-x)\left\{
	1+\frac{\alpha_s}{4\pi}C_F
					\left[\ln^2\left(\frac{1-x}{x}\right)
	+2-\frac{\pi^2}{3}\right]\right\}.
\end{eqnarray}
The transformation from a given scheme to a second one leads to a logarithmic
modification of the DA in order $\alpha_s$. In fact the problem arises
to which scheme the used non-perturbative input is related. Although in the
sum rule approach radiative corrections are considered as unimportant, for
this problem in question, it is interesting to study such perturbative
corrections. It is expected that radiative corrections are minimized in the
\mbox{$\mbox{CS}_{\rm I} $ } scheme.

Now we would like to discuss shortly the radiative corrections in their
dependence on $\omega$. Since $F_{\gamma\pi}(\omega,Q)$ is symmetric with
respect to $\omega\to -\omega$, it is sufficient to consider
$0\le\omega\le1$. From Eq.\ (\ref{TFF-Com-cOPE-fin}) it is obvious that for
$\omega <1$ the conformal partial waves for $2\le k$ are suppressed by
$\omega^k$. For $\omega=0$ only the first term contributes. The first moment
of the DA is uniquely given in terms of the pion decay constant; moreover,
the anomalous dimension $\gamma_0=0$, so the radiative correction, which is
also independent of the DA, is determined only by the $\alpha_s$ corrections
to $e_0$. In the conformal limit of the theory, we may conclude from Eq.\
(\ref{TFF-Com-cOPE-fin}) that the $\omega$ dependence of the perturbative
correction to the conformal partial waves is rather smooth.

\begin{figure}[htb]
\unitlength 1mm
\begin{center}
\begin{picture}(150,55)(0,-9)
\put(12,6){(a)}
\put(34,-6){$\textstyle Q^2\ ({\rm GeV}^2)$}
\put(0,0){\inclfig{15}{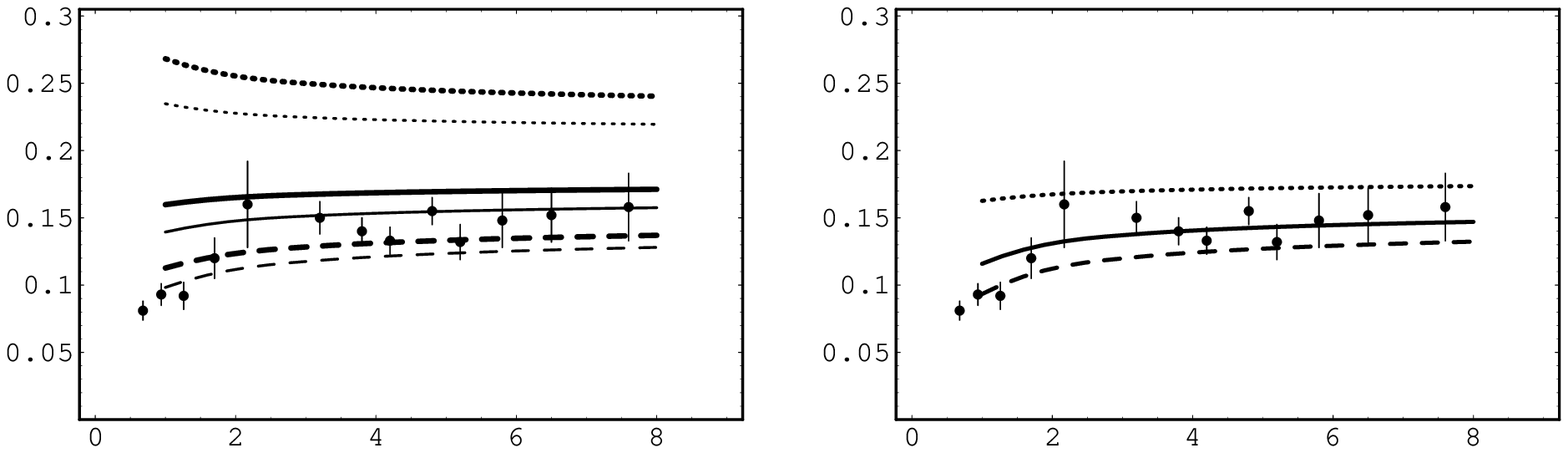}}
\put(92,6){(b)}
\put(114,-6){$\textstyle Q^2\ ({\rm GeV}^2)$}
\put(-5,8){\rotatebox{90}{$\textstyle Q^2 F_{\gamma\pi}(Q)\
								({\rm GeV}) $}}
\end{picture}
\end{center}
\caption[The SFA predictions for the pion to photon form factor to NLO.]{
The SFA predictions for the pion to photon form factor to NLO are shown for
the narrow (dashed line), asymptotic (solid line), and two-hump (dotted
line) DA, where the reference momentum is $Q_0=0.5$ GeV. In (a) the
predictions in the \mbox{ $\overline{\rm MS} $ } and \mbox{$\mbox{CS}_{\rm
I} $ } schemes are represented as thin and thick lines, respectively, where
$\Lambda^{\rm LO}=220$ MeV. In (b) the \mbox{$\mbox{CS}_{\rm II} $ } scheme
is chosen and $\Lambda^{\rm LO}=330$ MeV.
}
\label{Fig-TFF-NLO}
\end{figure}

In Fig.\ \ref{Fig-TFF-NLO}(a) the SFA predictions to NLO are compared to the
experimental data \cite{CELLO91,Sav95,Sav97} for the narrow, asymptotic, and
two-hump DA in both the \mbox{ $\overline{\rm MS} $ } and
\mbox{$\mbox{CS}_{\rm I} $ } scheme. The reference momentum for all DAs is
now set to $Q_0=0.5$ GeV and $\Lambda^{\rm LO}=220$ MeV. As pointed out
above the corrections in the conformal subtraction scheme
\mbox{$\mbox{CS}_{\rm I} $ } are about $10\%$ smaller than in the
$\overline{\rm MS}$ one. Caused by the NLO corrections the prediction of the
asymptotic DA now agrees better with the data, while the prediction of the
narrow DA starts to be below the data. In both schemes the two-hump DA
remains incompatible with the data. However, it can be demonstrated that its
prediction can be pushed down for a suitable choice of parameters. For
instance, if we choose the \mbox{$\mbox{CS}_{\rm II} $ } scheme and
$\Lambda^{\rm LO}=330$ MeV the predictions of all three DAs start to be
compatible with the data as illustrated in Fig.\ \ref{Fig-TFF-NLO}. This is
qualitatively the same effect as observed in Ref.\ \cite{CaoHuaMa96} in the
context of the model dependent study of the transverse momentum dependence.
Since this scheme transformation does not look very natural, we cannot draw
any conclusions, however, we may get some motivation to study perturbative
corrections beyond the NLO. Also we would like to note again that as
discussed above, the differences in the predictions arise mainly from the
fact that we "forget" to transform the DA.

\subsection{Beyond NLO}

As we can imagine from the results given in the previous subsection, for the
phenomenology of the considered transition form factor it is an important
task to study the corrections beyond the NLO. However, such calculations
have not been carried out at present. Fortunately, we can rely on the
conformal prediction (\ref{TFF-Com-cOPE-fin}), valid for $\beta=0$, where
the Wilson coefficients and the anomalous dimensions are known up to order
$\alpha_s^2$ from the results for the longitudinal structure function $g_1$
in DIS \cite{ZijNee94,CurFurPet80}. At NNLO it is known that the result
(\ref{TFF-Com-cOPE-fin-onshell}) is also spoiled in the
\mbox{$\mbox{CS}_{\rm I} $ } scheme by a term proportional to the
$\beta$-function \cite{GodKiv97,BelSch98}. For these restricted predictions,
symmetry breaking terms proportional to the $\beta$ coefficient are
neglected. Especially, for the asymptotic DA the NNNLO correction is
available from the Bjorken sum rule.

The photon-to-pion transition form factor predicted by the asymptotic DA is
given by the first term of the conformal expansion
(\ref{TFF-Com-cOPE-fin}):
\begin{eqnarray}
\label{TFF-Pre-conOPE}
Q^2 F_{\gamma\pi}(\omega,Q^2) &=& {\sqrt{2} f_\pi\over 3} 
	{2 \over 1+\omega}\,
	{_{2}F}_1\left({1,2 \atop 4}\Bigg|\frac{2\omega}{1+\omega}\right)
	c_0(\alpha_s).
\end{eqnarray} The coefficient
$c_0(\alpha_s)$ is normalized to 1 at LO. For the case that one
photon is almost real, i.e. $\omega=1$, we get 
\begin{eqnarray}
\label{TFF-Pre-conOPE-asy}
Q^2 F(1,Q^2) = \sqrt{2} f_\pi c_0(\alpha_s) =
0.185\,  c_0(\alpha_s) \mbox{\ GeV}.
\end{eqnarray}
In the conformal limit the first moment and thus also the asymptotic DA do
not evolve. The predictive power of the conformally covariant operator
product expansion (OPE) tells us that the coefficient $c_0(\alpha_s)$ is the
value of the Bjorken sum rule, which is calculated up to order $\alpha_s^3$
\cite{GorLar86,LarVer91}. For three active flavours the numerical result
reads\footnote{The $\alpha_s^4$-correction has been estimated to be
negative, too \cite{KatSta95,SamEllKar95}.}
\begin{eqnarray}
\label{BjoSumRul}
c_0(\alpha_s)=
	1-\frac{\alpha_s}{\pi}-3.58333\left(\frac{\alpha_s}{\pi}\right)^2-
	20.21527 \left(\frac{\alpha_s}{\pi}\right)^3+O\left(\alpha_s^4\right).
\end{eqnarray}
The higher-loop corrections at a scale of $Q^2=2\mbox{$\mbox{\ GeV}^2$}$,
where $\Lambda^{\rm LO}$ is assumed again to be $220$ MeV, reduce the LO
prediction to about 17\% in NNLO and to about 20\% in NNNLO (the NLO
contribution is 12\%).

From the size of the evolution effects arising in NLO we suspect that the
NNLO corrections remain small and can be neglected in a first step. To
obtain the NNLO corrections to the hard-scattering part for general DAs,
Eq.\ (\ref{TFF-Com-cOPE-fin-onshell}) is expanded up to order $\alpha_s^2$.
In the case of the two-hump DA (again evolved to a scale of $Q=\sqrt{2}$
GeV) the NNLO correction remains negligibly small, namely about -2\%. For
the chosen narrow DA the correction decreases from about -24\% in NLO to
about -30\% in NNLO.

We may conclude that in the conformal limit of the theory the perturbative
series looks very reasonable for DAs that are not ruled out by the data. It
would be very interesting to calculate the symmetry breaking effects in
NNLO, which can be done by taking into account only the $n_f$ dependent part
of the gluon vacuum polarization, which arise from the quark loop. Indeed,
this calculation has been carried out recently in the \mbox{ $\overline{\rm
MS} $ } scheme for the hard scattering part \cite{GodKiv97,BelSch98},
however, the given representation does not allow to transform this result to
the $\mbox{$\mbox{CS}_{\rm I} $ }$ scheme in a closed form. Moreover, to be
sure that the $n_f$ dependent term belongs to a conformal anomaly
proportional to the $\beta_0$ term, we have to combine the $n_f$ part of the
hard-scattering amplitude with that of the special conformal anomaly to NLO.
If one had at hand the desired corrections to the conformal prediction
(\ref{TFF-Com-cOPE-fin}), one would be able to fix the scale by the BLM
scale setting prescription.

\section{Pion form factor}
\label{FF}

The space-like elastic pion form factor has been extracted for $0.25\le Q^2
\le 6$\mbox{$\mbox{\ GeV}^2$}\ from the measured cross section of the
process $\gamma^\ast p\to \pi^+ n$. The intermediate off mass-shell pion was
extrapolated to the pion pole \cite{Beb78,Ame86}. Since this procedure
suffers from large systematic errors \cite{CarMil90}, a direct measurement
of the form factor would be very desirable. Indeed, the debate concerning
the applicability of the SFA to the electromagnetic pion form factors is
based on these data, which are doubted in Refs.\
\cite{CarMil90,BroRJiPanRob97}. In accordance with the counting rule, the
data behave as $1/Q^2$ for $Q^2 \ge 1$ \mbox{$\mbox{\ GeV}^2$}. In the SFA
the form factor is predicted to LO as:
\begin{eqnarray}
\label{pi-for-fac}
F_\pi(Q)=
	\frac{2\pi f_\pi^2 C_F \alpha_s(Q)}{3 Q^2} I_0(Q)^2, \quad
I_0(Q)=\int_0^1 dx\, \frac{\varphi(x,Q)}{1-x}.
\end{eqnarray}

\begin{figure}[htb]
\unitlength1mm
\begin{center}
\begin{picture}(90,57)
\put(6,7){\inclfig{7}{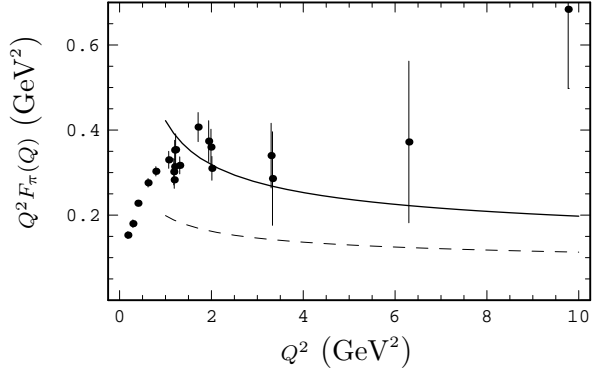}}
\put(0,18){\rotatebox{90}{
			$\scriptstyle Q^2 F_{\pi}(Q)\mbox{\ }
									  \big(\mbox{\small GeV}^2\big)$}}
\put(36,2){$\scriptstyle Q^2\mbox{\ } \big(\mbox{\small GeV}^2\big)$}
\end{picture}
\end{center}
\caption[Prediction for the electromagnetic pion form factor to leading
          order.]{
Predictions for the electromagnetic pion form factor at large momentum
transfer $Q$ for the asymptotic DA (dashed line) and the two-hump DA
(solid line) to LO, where $\Lambda_{\rm QCD} = 220$ MeV. 
Data are taken from Ref.\ \cite{Beb78} and references therein.}
\label{Fig-FF-Dat}
\end{figure}
\noindent
To fit the existing data shown in Fig.\ \ref{Fig-FF-Dat}, the value of $I_0$
should be about 4-5, so that an end-point enhanced DA such as the
two-hump function seems to be preferred.

The general factorization formula  for the electromagnetic pion form factor
at large momentum transfer reads
\begin{eqnarray}
F_{\pi}(Q)=
 \phi(x,\mu_F,\mu_R)\otimes T(x,y,Q,\mu_F,\mu_R) \otimes \phi(y,\mu_F,\mu_R),
\end{eqnarray}
where the scales are distinguished as the factorization scale $\mu_F$ and
the renormalization scale $\mu_R$. Note that also the DA depends on the
renormalization scale $\mu_R$. However, for simplicity we will not discuss
this dependence in detail and take into account evolution effects by the
solution of Eq.\ (\ref{evo-equ-mes}), in which both scales are identified.
The $\alpha_s$ correction to the hard-scattering part
\begin{eqnarray}
\label{piofor-TH}
T(x,y,Q^2)=
	{16\pi C_F \alpha_s(\mu_R)\over x y Q^2}\left[1+ 
		{\alpha_s(\mu_R)\over 2\pi} T^{(1)}(x,y,Q,\mu_F,\mu_R) +
		O\left(\alpha_s^2\right)\right]
\end{eqnarray}
was calculated by different authors in dimensional regularization; however,
different renormalizaton and factorization prescriptions were applied
\cite{FieGupOttCha81,DitRad81,Dit82,Sar82,RadKha85,BraTse87,MelNizPas98}.
The occurring differences are clarified in detail in Refs.\
\cite{RadKha85,BraTse87}. Indeed, if one takes into account the errors,
which are pointed out in Ref.\ \cite{BraTse87}, the results given in Refs.\
\cite{DitRad81,BraTse87} can be obtained from those in Refs.\
\cite{FieGupOttCha81,Dit82,Sar82,RadKha85,MelNizPas98}. The latter results
are based on the common renormalization operation and the prescription also
used in the calculation of the evolution kernel (a detailed discussion about
these items is given in Ref.\ \cite{RadKha85}):
\begin{eqnarray}
\label{piofor-THNLO}
T^{(1)}&=& C_F T^F(x,y,Q,\mu_F) + \beta_0 T^\beta(x,y,Q,\mu_R) + 
			  \left(C_F-C_A/2\right) T^{FA}(x,y),
\nonumber\\
 T^F&=&
			\left[3 + \ln(x y)\right] \ln\left(Q^2\over \mu^2_F\right) +
			\frac{1}{2} \ln^2(x y) + 3\ln(x y)-
            \frac{\ln x}{2(1-x)}- \frac{\ln y}{2(1-y)} -
			\frac{14}{3},
\nonumber\\
T^\beta&=&
	-\frac{1}{2}\ln\left(\frac{Q^2}{\mu_R^2}\right)-\frac{1}{2} \ln(x y)
	+\frac{5}{6},
\\
T^{FA}&=&
	{\rm Li}_2(1-x)-{\rm Li}_2(x) +
	\ln(1-x)\ln\left(\frac{y}{1-y}\right) -\frac{5}{3}
\nonumber \\
	&&+	\frac{1}{(x - y)^2}\Bigg(
	 (x+y-2x y)\ln(1-x) +2x y \ln(x)+\frac{(1-x)x^2+(1-y)y^2}{x-y}
\nonumber\\
	&&\times
\left[\ln(1-x)\ln(y) - {\rm Li}_2(1-x)+{\rm Li}_2(x)\right]
\Bigg) + \{x \leftrightarrow y\}.
\nonumber
\end{eqnarray}
The appearing $\ln x$ and $\ln y$ terms are responsible for large
contributions arising from the end-point region. Note that the singular
end-point behaviour of $T^{FA}(x,y)$ is $-\ln(x y)$, which is due to the
color structure suppressed by $1/N_c^2=1/9$. The appearing poles
in $(x-y)$ are actually cancelled by zeros, implying that the corresponding
term behaves smoothly for $x \to y$ (see Appendix \ref{App-TFA}).

In contrast to the photon-to-pion transition form factor we do not have
conformal predictions for the electromagnetic form factor that would enable
us to predict higher loop corrections without explicit calculation. However,
to analyse the NLO correction and to compare them in different schemes it is
possible to employ the conformal spin expansion:
\begin{eqnarray}
\label{piofor-TH-Exp}
F_{\pi}(Q^2)&=&
		{2\pi f_\pi^2 C_F \alpha_s(\mu_R)\over 3 Q^2}
		\sum_{n,m=0}^\infty {^\prime}
		\frac{2(3+2m)}{(m+1)(m+2)}\frac{2(3+2n)}{(n+1)(n+2)}
\\
		&\times& \langle \pi(P)| O_{mm}(\mu_F)|0\rangle^{red}
		\left[1+ 
		{\alpha_s(\mu_R)\over 2\pi} T^{(1)}_{mn}(Q,\mu_F,\mu_R) +
		\cdots \right]
		\langle 0| O_{nn}(\mu_F)|\pi(P)\rangle^{red},
\nonumber
\end{eqnarray}
where the relative NLO correction is given by the moments:
\begin{eqnarray}
\label{piofor-THconExp}
T^{(1)}_{mn} =  4
	\int_0^1dx\int_0^1dy\, (1-x)
 C_m^{\frac{3}{2}}(2x-1)
	T^{(1)}(x,y) (1-y) C_n^{\frac{3}{2}}(2y-1).
\end{eqnarray}
Using the shorthand notation
\begin{eqnarray}
\Big\langle\frac{f(x)}{x}\Big\rangle_m^{\rm rel}=
		\frac{\Big\langle\frac{f(x)}{x}\Big\rangle_m}
		{\Big\langle\frac{1}{x}\Big\rangle_m}, \quad
\Big\langle\frac{f(x)}{x}\Big\rangle_m =
\int_0^1 dx\, \frac{f(x)}{x}\frac{x(1-x)}{N_m}
 C_m^{\frac{3}{2}}(2x-1),
\end{eqnarray}
the result reads:
\begin{eqnarray}
\label{piofor-THNLO-Exp}
T_{mn}^{(1)}&=& C_F T^F_{mn} + \beta_0 T^\beta_{mn} + 
			  \left(C_F-C_A/2\right) T^{FA}_{mn},
\\
\label{piofor-THNLO-Exp-F}
 T^F_{mn} &=&
	\left[3+
	\Big\langle\frac{\ln x}{x}\Big\rangle_m^{\rm rel} +
	\Big\langle\frac{\ln x}{x}\Big\rangle_n^{\rm rel}
	\right] \ln\left(Q^2\over \mu^2_F\right) +
	\frac{1}{2} \Big\langle\frac{\ln^2 x}{x}\Big\rangle_m^{\rm rel} +
	\frac{1}{2} \Big\langle\frac{\ln^2 x}{x}\Big\rangle_n^{\rm rel}+
	\Big\langle\frac{\ln x}{x}\Big\rangle_m^{\rm rel} 
\nonumber\\
	&\times&\Big\langle\frac{\ln x}{x}\Big\rangle_n^{\rm rel}+
	\frac{5}{2} \Big\langle\frac{\ln x}{x}\Big\rangle_m^{\rm rel} +
	\frac{5}{2} \Big\langle\frac{\ln x}{x}\Big\rangle_n^{\rm rel} 
	-{\frac{14}{3}} -
	\frac{1}{2} \Big\langle\frac{\ln (1 - x)}{x}\Big\rangle_m^{\rm rel} -
   	\frac{1}{2} \Big\langle\frac{\ln (1 - x)}{x} \Big\rangle_n^{\rm rel},
\nonumber\\
{}
\\
\label{piofor-THNLO-Exp-beta}
T_{mn}^\beta &=&
	-\frac{1}{2}\ln\left(\frac{Q^2}{\mu_R^2}\right)-
	\frac{1}{2}\Big\langle\frac{\ln x}{x}\Big\rangle_m^{\rm rel} -
	\frac{1}{2}\Big\langle\frac{\ln x}{x}\Big\rangle_n^{\rm rel} 
	+\frac{5}{6},
\\
\label{piofor-THNLO-Exp-FA}
T_{mn}^{FA} &=&
		 - \Big\langle \frac{\ln x }{x} \Big\rangle_m^{\rm rel}  
      	\left(
				1 - \Big\langle \frac{\ln(1-x)+x}{x^2} \Big\rangle_n^{\rm rel}
		\right) -
		\Big\langle \frac{\ln x}{x} \Big\rangle_n^{\rm rel}  
      	\left(
		1 - \Big\langle \frac{\ln(1-x)+x}{x^2} \Big\rangle_m^{\rm rel}
		\right)
\nonumber\\
			&&+\frac{\pi^2}{3} - \frac{7}{3} + \Delta T^{FA}_{m} +
			\Delta T^{FA}_{n} +\Delta T^{FA}_{mn},
\end{eqnarray}
where $\Delta T^{FA}_{m}$ and $\Delta T^{FA}_{mn}$ are the (relative)
conformal moments of the functions $\Delta T^{FA}(x)$ and $\Delta
T^{FA}(x,y)$ defined in Eq.\ (\ref{new-rep-TFA}). The singular terms in Eq.\
(\ref{piofor-THNLO}), given by $\ln^i(x)/x$ for $i=0,1,2$, provide a
$\ln^i(m)$-behaviour and are analytically calculated in Appendix
\ref{ChapASB}. The remaining regular part provides $\Delta T^{FA}_{m}$,
$\Delta T^{FA}_{m}$ and $\Delta T^{FA}_{mn}$, which are power and $1/N_c^2$
suppressed. Therefore, for all DA this part will become negligibly small for
growing $m$ and $n$ and thus it is sufficient to calculate only the first
few moments. The numerical results are listed in Appendix \ref{ChapASB},
too.

For the chosen set of DA's it turns out that $T^F$ gives a moderate negative
contribution, while $T^\beta$ contains a large positive contribution. There
are different suggestions to optimize both the factorization and
renormalization scale setting. Here we will skip the factorization scale
setting problem (detailed discussions are given for instance in Refs.\
\cite{DitRad81,MelNizPas98,SteSchKim98}) and deal only with the
renormalization scale setting prescription. This contribution arises only
from the renormalization of the coupling and is proportional to $\beta_0$.
Thus, it can be absorbed in a quite natural way by the BLM prescription into
the scale of the coupling \cite{BroRJiPanRob97}:
\begin{eqnarray}
\mu_R \to Q^\ast_{mn}= Q \exp\left\{-
	\frac{1}{2}\Big\langle\frac{\ln x}{x}\Big\rangle_m^{\rm rel} -
	\frac{1}{2}\Big\langle\frac{\ln x}{x}\Big\rangle_n^{\rm rel} 
	+\frac{5}{6}\right\},
\end{eqnarray}
with $T^\beta_{m n}(Q,Q_{mn})$ vanishing. For the ``natural'' scale setting
prescription $\mu_F=Q$, the form factor can be written after summation as 
\begin{eqnarray}
\label{FF-predic}
F_\pi(Q)&=&
	{2\pi C_F f_\pi^2 \alpha_s(Q^\ast)\over 3 Q^2}  \left\{1+ 
		{\alpha_s\over 2\pi} r^{(1)}(Q,Q) + O\left(\alpha_s^2\right)\right\}
		I(Q,Q_0)^2,
\\
&=& {\alpha_s(Q^\ast)\over Q^2} \left\{1+ 
		{\alpha_s\over 2\pi} r^{(1)}(Q,Q) +
	 O\left(\alpha_s^2\right)\right\} C_\pi(Q,Q_0),
\nonumber
\end{eqnarray}
where  $C_\pi(Q,Q_0)= 2\pi C_F f_\pi^2 I(Q,Q_0)^2/3$
contains all evolution effects, $Q^\ast=Q e^{-\Delta}$ and the $I$ integral
is defined in Eq.\ (\ref{I-integral}). To obtain the NLO corrections as
scheme independent as possible, one should take into account the higher order
corrections due to the evolution in the following manner:
\begin{eqnarray}
C_\pi(Q,Q_0)=C^{\rm LO}_\pi(Q,Q_0)\left(
		1+\frac{\alpha_s}{2\pi}\cdots + O(\alpha_s^2)\right)
\end{eqnarray}
and truncate then the series in Eq.\ (\ref{FF-predic}) in the first order in
$\alpha_s$. Fortunately, for a reference momentum square of $Q\ge 2$ GeV the
size of the NLO contributions arising from the evolution allows us to use
Eq.\ (\ref{FF-predic}) in practice. In Table \ref{tab-FF-HS} we give the
results for $C_\pi(Q_0,Q_0)$, $r^{(1)}$ and $\Delta$ at
$Q^2=2$\mbox{$\mbox{\ GeV}^2$}\ for the chosen set of DAs and for the three
considered schemes. To carry out the calculation more easily, the scale
setting has been done after summation.
\begin{table}[htb]
\begin{center}
\begin{tabular}{|c||c||c|c||c|c||c|c|}\hline
DA  &\mbox{\ } &\multicolumn{2}{c||}{$\overline{\rm MS}$}
&\multicolumn{2}{c||}{$\mbox{CS}_{\rm I}$}
&\multicolumn{2}{c|}{$\mbox{CS}_{\rm II}$}\\
\mbox{\ }  &$C_\pi$
&\multicolumn{1}{c|}{$r^{(1)}$ }&\multicolumn{1}{c||}{$\Delta$ }
&\multicolumn{1}{c|}{$r^{(1)}$ }&\multicolumn{1}{c||}{$\Delta$ }
&\multicolumn{1}{c|}{$r^{(1)}$ }&\multicolumn{1}{c|}{$\Delta$ }\\
\hline\hline
$\varphi^{10}$ 	&0.211	&-8.1&1.60	&-5.7&1.60	&-5.7&1.42	 \\\hline
$\varphi^{\rm as}$	&0.431	&-7.8&2.33	&-5.2&2.33	&-5.2&2.00  \\\hline
$\varphi^{1/2} $ 	&0.767	&-1.22&3.22	&0.82&3.22  &0.82&2.62   \\\hline
$\varphi^{\rm CZ}$	&0.848	&-5.9&2.93	&-3.1&2.93  &-3.1&2.46   \\\hline
$\varphi^{1/4} $ 	&1.725	&34.4  &5.10 &29.8&5.10  	&29.8&3.8   \\\hline
\end{tabular}
\end{center}
\caption[Predictions for the hard-scattering amplitude
of the electromagnetic pion form factor in LO and NLO.]{
LO  predictions and NLO corrections for the hard-scattering amplitude
of the electromagnetic pion form factor for different DAs, 
where the evolution is neglected and the factorization scale is set to
$Q$. The renormalization scale is fixed by the BLM procedure.}
\label{tab-FF-HS}
\end{table}

Finally, we discuss the phenomenological consequences of the perturbative
NLO corrections. Due to the large value of $\Delta$ the scale $Q^\ast$
appearing in the coupling will be very small. For instance, at $Q=\sqrt{2}$
GeV we find for the asymptotic DA $Q^\ast\approx 0.14$ GeV and for the
two-hump DA $Q^\ast\approx 0.08$ GeV in the \mbox{ $\overline{\rm MS} $ }
and \mbox{$\mbox{CS}_{\rm I} $ } schemes. At this low scale the perturbative
treatment of the coupling is not valid anymore. However, it was argued that
the coupling is frozen at low momentum transfer. A simple parametrization
of such a frozen coupling arises from the idea that the gluon propagator has
an effective mass $m_g$ that is induced by non-perturbative effects
\cite{Cor82,DonLan89,DucHalNat93}:
\begin{eqnarray}
\alpha_s^{\rm fro}(Q)=
 \frac{4\pi}{\beta_0 \ln\left(\frac{Q^2+4 m_g^2}{\Lambda^2}\right)}.
\end{eqnarray}
In the following we assume that $\alpha_s^{\rm fro}$ is about 0.5 at very
low momentum transfer \cite{MatSte94,BroRJiPanRob97}. Employing Eq.\
(\ref{FF-predic}) we obtain from Table \ref{tab-FF-HS} the NLO prediction
for the electromagnetic form factor. In the conformal scheme
\mbox{$\mbox{CS}_{\rm I} $ } the prediction is $Q^2 F_\pi \approx 0.12$ for
the asymptotic DA and $Q^2 F_\pi \approx 0.32$ for the two-hump DA. These
results are compatible with the LO predictions shown in Fig.\
\ref{Fig-FF-Dat}.

\section{Conclusions}
\label{Concl}

In this paper we have studied the scheme dependence of the photon-to-pion
transition form factor and the electromagnetic pion form factor in NLO for
two different schemes, namely, the popular $\overline{\mbox{MS}}$ scheme and
the so-called conformal scheme. In the case of two-particle distribution
amplitudes, the latter one is uniquely defined in the conformal limit of the
theory by the requirement of conformal covariance. Obviously, beyond this
limit we have to deal with an ambiguity that is proportional to the $\beta$
function. Such terms can be naturally included in the scale of the coupling
by the BLM scale setting prescription. However, as we saw in Subsection
\ref{NLO-DA-B} in the case of exclusive processes this prescription is not
sufficient to restore the conformal covariance of the perturbative
prediction with respect to the original representation. We may expect that
for so-called reducible but non-decomposable representations conformal
covariance holds true.

Instead of dealing with the usual convolution of hard scattering part and
DAs we exploit the possibility to work directly with the conformal moments,
so that the QCD prediction for exclusive processes is given as a sum about
such moments. The advantage of such representation is that the evolution of
the moments is easy to handle and the back transformation into the $x$-space
by infinite sums over Gegenbauer polynomials can be avoided. Although such
sums can be calculated with an appropriate accuracy in a straightforward
way, the numerical cancellation of the oscillations requires some care. For
the method, which we used here, the exclusive predictions by themselfs are
given by infinite sums that do not suffer under oscillations and so they can
be calculated without difficulties by taking into account a sufficient large
number of terms. Especially, for the photon-to-pion transition form factor
this method allows us to immediately use  information available from DIS to
give a prediction beyond the NLO.

The NLO corrections in the conformal scheme \mbox{$\mbox{CS}_{\rm I} $ } are
in general smaller than in the \mbox{ $\overline{\rm MS} $ } scheme. This
observation supports the general argument that one should choose a scheme in
which the underlying symmetries of the theory are preserved in the maximal
possible manner. In the case of conformal symmetry we cannot fix the scheme
uniquely, since the renormalization of the coupling causes a conformal
anomaly proportional to the $\beta$ function. As we demonstrated this
remaining freedom can provide us huge differences for the NLO corrections.
In the case of the photon-to-pion transition form factor we saw that due to
this freedom the differences between the prediction of end-point narrow and
concentrated DAs are starting to be compatible with the data. However, one
has to be very careful with some conclusions. Since we did not transform the
DAs, the real question arising here is: To which scheme belong the
non-perturbative input DAs? It seems to us that this problem should be
considered with more attention.

It is also desirable to go beyond the NLO to get more insight in the
perturbative corrections to exclusive processes. At least for the
photon-to-pion transition form factor the conformal techniques can be
extended for the calculation of the NNLO corrections to the coefficient
function in the full theory.

\appendix
\section{Decomposition of $T^{FA}$}
\label{App-TFA}

In this appendix we extract from $T^{FA}(x,y)$, defined in Eq.\
(\ref{piofor-THNLO}), that part which vanishes in the limit $x,y \to 0$.
Moreover, we derive for these terms an expansion which can be conveniently
used in the convolution of the hard-scattering amplitude with the DA's. The
most complicated looking term in $T^{FA}(x,y)$ is proportional to
$1/(x-y)^2$:
\begin{eqnarray}
t^{FA}&=&
	\frac{1}{(x - y)^2}\Bigg(
	 (x+y-2x y)\ln(1-x) +2x y \ln(x)+\frac{(1-x)x^2+(1-y)y^2}{x-y}
\\
	&&\times
\left[\ln(1-x)\ln(y) - {\rm Li}_2(1-x)+{\rm Li}_2(x)\right]
\Bigg) + \{x \leftrightarrow y\}.
\nonumber
\end{eqnarray}
Using the simple algebraic decomposition
\begin{eqnarray}
\frac{(1-x)x^2 +(1-y) y^2}{(x-y)^3} =
\frac{1-x-y}{x-y} + \frac{(2-x-y)x y}{(x-y)^3}
\end{eqnarray}
and the equality
\begin{eqnarray}
\label{idenLi}
{\rm Li}_2(1-x) = -{\rm Li}_2(x) - \ln(1-x) \ln x + \frac{\pi^2}{6},
\end{eqnarray}
we immediately obtain for $t^{FA}$:
\begin{eqnarray}
t^{FA}&=&
\frac{1}{x-y}\Bigg[
 \left\{1+(1-x-y)\ln(x y)\right\}\ln\left(\frac{1-x}{1-y}\right) +
2(1-x-y) \left\{{\rm Li}_2(x)-{\rm Li}_2(y)\right\}\Bigg]
\nonumber \\
&&+ 
	\frac{\ln(x y)}{(x - y)^2}\Bigg[
	 2x y +\frac{(2-x-y)x y}{x-y} \ln\left(\frac{1-x}{1-y}\right)
\Bigg] + \frac{1}{(x - y)^2}\Bigg[
	 (x+y-2x y)
\\
&&\times \ln\left[(1-x)(1-y)\right]+\frac{2(2-x-y)x y}{x-y} 
\left[{\rm Li}_2(x)- {\rm Li}_2(y)\right]-
(x-y)\ln\left(\frac{1-x}{1-y}\right) \Bigg] .
\nonumber
\end{eqnarray}
To extract the desired part, one should take into account the
following set of algebraic identities:
\begin{eqnarray}
\frac{f(x)-f(y)}{x-y} &=&
-1 + \frac{x+f(x)}{x} + \frac{y+f(y)}{y}+
\frac{y^2 f(x)- (x-y)x y - x^2 f(y)}{(x-y)x y},
\nonumber \\
\frac{(1-x-y)[f(x)-f(y)]}{x-y} &=&
\mp 1 + \frac{\pm x+(1-x) f(x)}{x}+ \frac{\pm y+(1-y)f(y)}{y}
\\
&&
+\frac{(1-2x)y^2 f(x)\mp(x-y)x y - (1-2y) x^2 f(y)}{(x-y)x y},
\nonumber
\end{eqnarray}
where in the last equation the upper [lower] sign is used for $f(x)=
\ln(1-x) \left[{\rm Li}_2(x)\right]$. Note that the third term on the
RHS vanishes for $x,y \to 0$. Taking also into account the remaining
part of $T^{FA}(x,y)$, we finally obtain after some simple aglebra the
desired representation:
\begin{eqnarray}
\label{new-rep-TFA}
T^{FA}(x,y)&=&
- \ln(x)\left[1-\frac{\ln(1-y)+y}{y}\right]-
\ln(y)\left[1-\frac{\ln(1-x)+x}{x}\right] + \frac{\pi^2-7}{3}
\nonumber \\
&&+\Delta T^{FA}(x) + \Delta T^{FA}(y) + \Delta T^{FA}(x,y),
\nonumber\\
\Delta T^{FA}(x)&=&
	\frac{\ln(1-x)+x}{x} +\frac{\ln(x)\left[(1-2x)\ln(1-x)+x\right]}{x}
	+2\frac{(1-2x){\rm Li}_2(x)-x}{x},
\nonumber\\
\Delta T^{FA}(x,y)&=&
-2 \ln(1-x)\ln(1-y)+ \theta^{FA}(x,y),
\nonumber\\
\theta^{FA}(x,y)&=& 
\frac{ \ln(x y)}{(x-y)x y} \left[
	y^2	(1-2x) \ln(1-x) -x y (x-y) -x^2 (1-2y) \ln(1-y)
							\right]
\\
&&+ \frac{1}{(x-y)x y} \Bigg[
		y^2 \left\{2(1-2x){\rm Li}_2(x) + \ln(1-x)\right\}
+x y (x-y)
\nonumber \\
&&  \hspace{2.5cm} -x^2 \left\{2(1-2y){\rm Li}_2(y) + \ln(1-y)\right\}
	- x y \ln\left(\frac{1-x}{1-y}\right)				\Bigg]
\nonumber \\
&&+ 	\frac{\ln(x y)}{(x - y)^2}\Bigg[
	 2x y +\frac{(2-x-y)x y}{x-y} \ln\left(\frac{1-x}{1-y}\right)
\Bigg] + \frac{1}{(x - y)^2}
\nonumber\\
&&\times\Bigg[
	 (x+y-2x y) \ln\left[(1-x)(1-y)\right]+\frac{2(2-x-y)x y}{x-y} 
\left[{\rm Li}_2(x)- {\rm Li}_2(y)\right]\Bigg] ,
\nonumber
\end{eqnarray}
where $\Delta T^{FA}(x)$ and $\Delta T^{FA}(x,y)$
vanish for $x,y \to 0$. Especially, for $\Delta T^{FA}(x,y)$
this can be seen by a power expansion of $\theta^{FA}(x,y)$, namely,
\begin{eqnarray}
\label{def-theta}
\theta^{FA} &=&
 x y \sum_{i=0}^{\infty} \sum_{j=0}^i
			\left(
 \frac{3-i j +j^2}{(2+i)(3+i)} 	\ln (x y) -
	\frac{2(1-i-i j+j^2)}{(2+i)^2} -\frac{3+3 i+2 i j-2 j^2}{(3+i)^2}
\right) x^j y^{i-j}.
\nonumber\\
{\ }
\end{eqnarray}
It is now also obvious that $\theta^{FA}(x,y)$ is regular at $x=y$. In the
calculation of the NLO corrections it is justified to approximate
$\theta^{FA}(x,y)$ by the first few terms of this expansion.

\section{Conformal moments}
\label{ChapASB}

Here we list the needed conformal expansions (with respect to the
Gegenbauer polynomials) of terms appearing in the hard-scattering amplitude
of the pion form factor to NLO. They are obtained by employing the
representation
\begin{eqnarray}
\frac{x(1-x)}{N_k} C_k^{\frac{3}{2}}(2x-1)=
	(-1)^k \frac{2(3+2k)}{(k+1)!} \frac{d^k}{dx^k} x^{k+1}(1-x)^{k+1}
\end{eqnarray}
as well as the definition of the $B$-function
\begin{eqnarray}
B(a,b)=
	\frac{\Gamma(a)\Gamma(b)}{\Gamma(a+b)}=
	\int_0^1 dx\, x^{a-1}(1-x)^{b-1}.
\end{eqnarray}

\begin{eqnarray}
\Big\langle \frac{\ln^i x}{x}\Big\rangle_k &=&
	\frac{\partial^i}{\partial\epsilon^i}
	\int_0^1 dx\ x^{\epsilon-1} \frac{x(1-x)}{N_k}
							C_k^{\frac{3}{2}}(2x-1)_{|\epsilon=0}
\nonumber\\
	&=&\frac{\partial^i}{\partial\epsilon^i}\frac{2(3+2k)}{(k+1)!}
	\int_0^1 dx\,\left[\frac{d^k}{dx^k}
			x^{\epsilon-1}\right] x^{k+1}(1-x)^{k+1}_{|\epsilon=0}
\\
	&=&(-1)^k \frac{\partial^i}{\partial\epsilon^i} 
	\left[\frac{2(3+2k)\Gamma(1+\epsilon)\Gamma(k+1-\epsilon)}
	{\Gamma(1-\epsilon)\Gamma(k+3+\epsilon)}\right]_{|\epsilon=0}.
\nonumber			
\end{eqnarray}
For $i=0$ we immediately find
\begin{eqnarray}
\Big\langle \frac{1}{x}\Big\rangle_k &=&
 (-1)^k
	\frac{2(3+2k)}{(k+1)(k+2)}.
\end{eqnarray}
Applying the definition of the digamma function
$\psi(x)= \frac{d}{dx}\ln\Gamma(x)$ provides
\begin{eqnarray}
\Big\langle \frac{\ln x}{x}\Big\rangle_k &=&
 (-1)^k
	\frac{2(3+2k)}{(k+1)(k+2)}\left[2\psi(1)-\psi(k+1)-\psi(k+3)\right],
\\
\Big\langle \frac{\ln^2 x}{x}\Big\rangle_k &=&
 (-1)^k
	\frac{2(3+2k)}{(k+1)(k+2)}
			\bigg\{
	[2\psi(1)-\psi(k+1)-\psi(k+3)]^2
\\
&&\hspace{5.5cm}+\psi'(k+1)-\psi'(k+3)\bigg\}.
\nonumber
\end{eqnarray}
For large $k$ we find with $\psi(k+1)-\psi(1)=\ln(k+1) + O(1/k)$
the following asymptotic behaviour
\begin{eqnarray}
\Big\langle \frac{\ln^i x} {x}\Big\rangle_k =
	\Big\langle \frac{1}{x}\Big\rangle_k \left\{\left[2\ln(k+1)\right]^i+
			O(1/k)\right\},
\end{eqnarray}
which reflects the singular end-point behaviour for $x\to 0$.

Closed formulas for $\langle \frac{\ln(1-x)}{x}\rangle_k$ and
$\langle \frac{\ln(1-x)+x}{x^2}\rangle_k$ can be derived
by employing the expansion
$\frac{\ln(1-x)}{x}=-\sum_{i=0}^\infty x^i/(i+1)$. After integration over $x$
the summation can be carried out:
\begin{eqnarray}
\Big\langle \frac{\ln(1-x)}{x}\Big\rangle_k =
-\frac{2(3+2k)}{(k+2)!} \sum_{n=k}^{\infty} \frac{n!}{(n-k)!} B(n+1,k+3)
= -\frac{2(3+2k)}{(k+1)^2 (k+2)^2}.
\end{eqnarray}
In an analogous way one finds:
\begin{eqnarray}
\Big\langle \frac{\ln(1-x)+x}{x^2}\Big\rangle_k &=&
-\frac{2(3+2k)}{(k+1)!} \sum_{n=k}^{\infty} \frac{1}{n+2}\frac{n!}{(n-k)!}
B(n+2,k+2)
\nonumber\\
&=& -\frac{2(3+2k)}{(k+1)(k+2)^2 (k+3)}
{_3F_2}\left({k+2, 1, 2 \atop k+3,k+4 }\Bigg|1\right).
\end{eqnarray}

Finally, the numerical results that are needed for the calculation
of the pion form factor to NLO will be listed in Tables \ref{tab-num1-HS}
and \ref{tab-num2-HS}.
\begin{table}[htb]
\begin{center}
\begin{tabular}{|c||c|c|c|c|c|c|c|c|c|}\hline
$m$ & 0 & 2 & 4 & 6 & 8 & 10 & 12 & 14
\\ \hline\hline
$\Big\langle \frac{\ln(1-x)+x}{x^2}\Big\rangle_m^{\rm rel}$ & -0.290 & -0.073
& -0.031 & -0.017 & -0.011 & -0.007  & -0.005 & -0.004
\\ \hline\hline
$\Delta T^{FA}_m$ &  -2.157 & -0.226 & -0.086 & -0.045 & -0.028 & -0.019
& -0.013 & -0.010
\\ \hline
\end{tabular}
\end{center}
\caption[The first few moments of
	$\Big\langle \frac{\ln(1-x)+x}{x^2}\Big\rangle_m^{\rm rel}$ and
	$\Delta T^{FA}_m$.]
{Numerical results for the first few moments of
	$\Big\langle \frac{\ln(1-x)+x}{x^2}\Big\rangle_m^{\rm rel}$ and
	$\Delta T^{FA}_m$. The latter is the relative conformal moment
of $\Delta T^{FA}(x)$, defined in Eq.\ \ref{new-rep-TFA}, and
appears in Eq.\ (\ref{piofor-THNLO-Exp-FA})}
\label{tab-num1-HS}
\end{table}
\begin{table}[htb]
\label{tab4}
\begin{center}
\begin{tabular}{|c||c|c|c|c|c|c|}\hline
${\mbox{\ }\mbox{\ }m \atop  n\mbox{\ }} $ & 0 & 2 & 4 & 6 & 8 & 10
														\\\hline\hline
 0 &-0.845 & -0.129 & -0.050 & -0.026 & -0.016 & -0.010 \\ \hline
 2 &-0.129 & -0.020 & -0.008 & -0.005 & -0.003 & -0.002 \\ \hline
 4 & -0.050 & -0.008 & -0.004 & -0.002 & -0.002 & -0.001 \\ \hline
 6 & -0.026 & -0.005 & -0.002 & -0.002 & -0.001 & -0.001 \\ \hline
 8 &-0.016 & -0.003 & -0.002 & -0.001 & -0.001 & -0.001 \\ \hline
10 & -0.010 & -0.002 & -0.001 & -0.001 & -0.001 & 0.000 \\ \hline
\end{tabular}
\end{center}
\caption[The first few moments of $\Delta T^{FA}_{mn}.$]
{Numerical results for the first few relative conformal moment
of $\Delta T^{FA}(x,y)$ defined in Eq.\ \ref{new-rep-TFA}. They appear in
Eq.\ (\ref{piofor-THNLO-Exp-FA}).}
\label{tab-num2-HS}
\end{table}
\noindent
Let us remark that the moments $\Delta T^{FA}_{mn}$ can be easily
calculated by employing the representation (\ref{def-theta}).


\end{document}